# Complex chiral columns made of achiral quinoxaline derivatives with semi-flexible cores

Paulina Rybak*, Adam Krowczynski, Jadwiga Szydlowska, Damian Pociecha and Ewa Gorecka

Department of Chemistry, University of Warsaw, ul. Zwirki i Wigury 101, 02-089 Warsaw (Poland)

**ABSTRACT:** Mesogenic materials, quinoxaline derivatives with semi-flexible cores, are reported to form new type of 3D columnar structure with large crystallographic unit cell and *Fddd* symmetry below columnar hexagonal phase. The 3D columnar structure is a result of frustration imposed by arrangement of helical columns of opposite chirality into triangular lattice. The studied materials exhibit fluorescent properties that could be easily tuned by modification of molecular structure, compounds with the extended π electron conjugated systems form aggregates and fluorescence is quenched. For molecules with flexible structure the fluorescence quantum yield reaches 25%. On the other hand, compounds with more rigid mesogenic core, for which fluorescence is suppressed show strong hole photocurrent. For some materials also bi-polar: hole and electron transfer was observed.

## Introduction

In recent years, columnar liquid crystalline phases that are usually formed by flat and rigid molecules having at least a few side alkyl chains have attracted great interest from both scientific and industrial point of view. They have been tested for applications in various areas, such as optical device construction [1] and in solar cells [2-5]. An advantage of organic materials for such applications, as opposed to their inorganic counterparts, is that they are soft, and relatively easy to align, and therefore susceptible to external manipulation. In electronic applications often high and anisotropic charge carrier mobility is required; in columnar structure this condition is easily fulfill as each column provides a 1D path for charge transport through its central part made of mesogenic cores, while the alkyl chains isolate those inner "wires" [6,7]. Many types of molecules with an extended system of aromatic rings in mesogenic core have been tested for the design of optoelectronic devices such as OLEDs (organic light-emitting diodes) [8-10], OFETs (organic field-effect transistors) [11-13] and other photovoltaic applications [14-16], however the performance of such materials is still problematic. The requirements – e.g. good charge transport and strong florescence for OLEDs, are often contradictory. They could be met by designing molecules with some conformational freedom; molecular flexibility may promote fluorescence by preventing formation of H-aggregates [17,18], while π - π interactions confined to certain molecular fragments may enhance charge transfer. Quinoxaline derivatives are perfect candidates for such materials, from an organic synthesis point of view they are easy to modify, and if properly functionalized they are able to generate a columnar packing [19-21]. Quinoxaline derivatives of lath-like shape may be stacked with random orientation leading to columns with overall circular cross-section, but might be also arranged in helical stacks if the neighboring molecules twist with respect to each other to provide more space for the alkyl chains. The packing of helical columns can be locally homochiral [22,23] but in general columns of opposite chirality can also alternate in the crystal lattice (racemic or antichiral structure)[24]. However, the regular antichiral structure cannot be realized if the columns are arranged in a hexagonal lattice, in this case the frustration should lead to some distortion of the structure, since an ideal triangular grid cannot be decorated evenly with two types of objects - the left and right handed helices[25]. How the helical columns are arranged into a frustrated pseudo-hexagonal structure (with *Fddd* symmetry) has been recently studied and discussed in relation to other systems built from achiral molecules.[26]

Here we report systematic studies of the relation between rigidity of the molecular structure and the phase behavior of quinoxaline based mesogenic molecules. Apart from columnar hexagonal phase, in two of the studied compounds we discovered additional three-dimensional LC phase built of helical columns.

## Results and discussion

We have synthesized several quinoxaline derivatives and. The chosen compounds **1-5** (*Figure 1*) exhibiting liquid crystalline phases in a broad temperature range, were studied in details. Other non-liquid-crystalline compounds with similar structures are presented in SI (*Figure S1*).

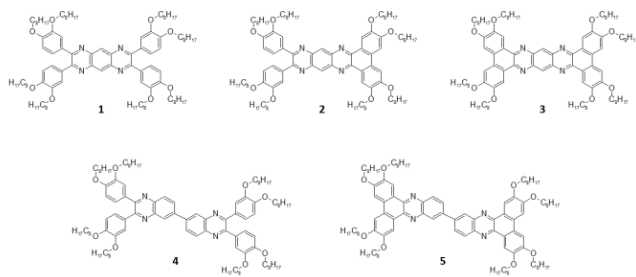

**Figure 1.** Chosen molecules that exhibited liquid crystalline properties under the isotropisation temperature.

The molecular structure was systematically modified to correlate its flexibility with phase properties. In molecules **1 - 3** the central core is rigid, made with three adjacent rings. In molecule **1** all four outer phenyl rings attached to pyrazino-quinoxaline center are free to rotate; in contrast, in asymmetric molecule **2** only two of them were incorporated into conjugated structure, thus their rotation is blocked. Finally in molecule **3** all four outer phenyl rings are in the conjugated structure of large mesogenic core. In molecules **4** and **5** the flexibility was introduced in the central part of the molecule - two quinoxaline units were joined by single bond; additionally the outer rings are free to rotate for molecule **4** while they are incorporated into conjugated system of mesogenic units for molecule **5**.

Observed phase sequences, phase transition temperatures with corresponding enthalpy changes for compounds **1 - 5** are presented in the *Table 1*. Compounds **1** [27] and **4** [28] were studied previously by other groups, however our findings on phase sequence for compound **1** differ from the former report.

**Table 1.** Phase transition temperatures [°C] and associated thermal effects [J/g] (in parentheses) for obtained compounds.

| No. | Phase sequence |
|---|---|
| 1 | **Cry** - 86.19 (33.70) –**Fddd** 89.42 (13.33) - **Col**$_h$ - 102.76 (6.52) - **Iso** |
| 2 | **Cry** - 73.77 (21.59) - **Col**$_h$ - 155.00 (1.40) - **Iso** |
| 3 | **Cry** - 122.79 (49.53) - **Col**$_h$ - 240.00 (2.65) - **Iso** |
| 4 | **Cry** - 58.65 (6.07) - **Col**$_h$ - 81.49 (2.51) - **Iso** |
| 5 | **Cry** - 135.20 (35.28) - **Fddd** - 189.06 (3.17) - **Col**$_h$ - 196.46 (6.91) - **Iso** |

Studied compounds **1-5** exhibit columnar hexagonal phase and their clearing temperatures depend strongly on the flexibility of the mesogenic core. The highest stability of liquid crystalline state was observed for compounds **3** and **5**, which molecules have fully restricted rotation of external phenyl rings. The hexagonal columnar phase observed between crossed polarizers exhibited either typical fan texture with extinction directions along polarizers or homeotropic texture in parts of the sample with columns axis oriented perpendicular to the surface (*Figure 2a*). Two of the materials, **1** and **5**, exhibited additional columnar phase below Col$_h$ phase. For compound **1** nearly no texture changes were observed at the transition temperature, while for material **5** the homeotropic areas became weakly birefringent in lower temperature phase (*Figure 2b*), three types of domains were observed that can be sequentially brought into extinction condition between crossed polarizers by rotation of the sample by 30 degrees (*Figure 2c-2e*). This shows that columns tilt with respect to the substrate surface in the directions enforced by hexagonal symmetry. Upon, heating the sample to Col$_h$ phase the domains disappear but the residual birefringence in previously homeotropic areas is still detectable, apparently the columns remain tilted to the surface.

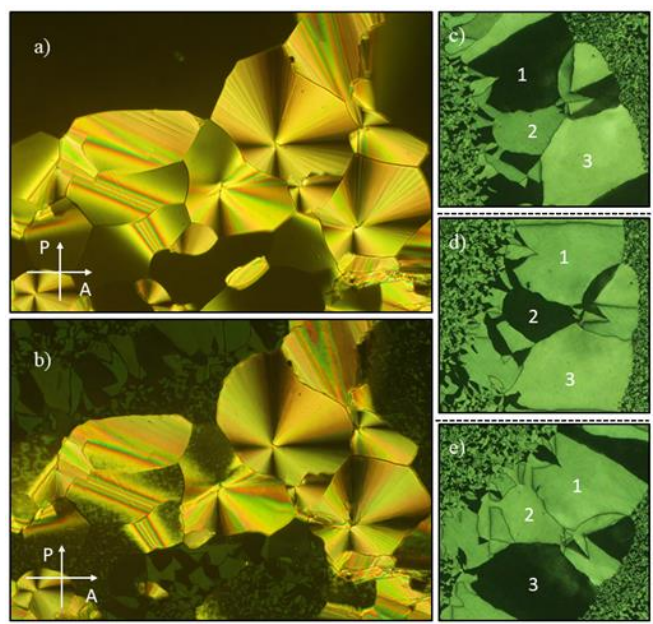

**Figure 2.** a) Microscopic texture of compound **5** at 195°C (*Col$_h$* phase) with visible homeotropic areas and b) at 170°C (*Fddd* phase); c), d) and e) in *Fddd* phase domains can be brought into extinction condition by rotating the sample by 30°.

**Table 2.** Crystallographic distances corresponding to X-ray diffraction signals, calculated parameters of crystallographic lattice, **a**, and distance between discs along the columns, **d**.

| No. | T/°C | Phase | signals/Å | a/Å | d/Å |
|---|---|---|---|---|---|
| 1 | 102 | Col$_h$ | 25.6; 14.8; 12.8 | 29.6 | 4.6 |
| 2 | 150 | Col$_h$ | 25.0; 12.6 | 29.0 | 4.5 |
| 3 | 160 | Col$_h$ | 24.4; 12.2 | 28.2 | 5.0 |
| 4 | 70 | Col$_h$ | 29.2; 16.8; 14.6 | 31.9 | 4.4 |
| 5 | 190 | Col$_h$ | 26.7 ; 15.6 ; 13.4 | 31.1 | 4.8 |

Phase assignment was confirmed using X-ray diffraction (XRD), the crystallographic parameters for studied materials were collected in *Table 2 and 3*.

In XRD patterns for columnar hexagonal phase the characteristic set of sharp low-angle signals was detected, with the peak positions in q-space being in ratio 1:√3:2. For the sample with one free surface, the columns were spontaneously aligned parallel to the sample surface but the azimuthal orientation of columns was degenerated. In result the x-ray pattern with 6-fold symmetry but unequal intensities of the signals at the same q position was registered (*Figure S3a*)

**Table 3.** Crystallographic unit cell parameters of the *Fddd* orthorhombic phase, extracted from X-ray diffraction signals.

| No. | a/Å | b/Å | c/Å |
|---|---|---|---|
| 1 | 112.2 | 64.8 | 156.6 |
| 5 | 107.1 | 61.8 | 50.9 |

In the high-angle range of the patterns the diffused reflection denotes short range positional order of molecules along the columns, with mean intermolecular distance ~4.5 Å (see SI, *Figure S3*). For compounds **1 – 3** containing the benzenetetraamine core, the hexagonal lattice parameter, reflecting the column diameter, was ~29 Å, for the ones containing diaminobenzidine (molecules **4** and **5**) it was ~31 Å; the difference is expected as diaminobenzidine derivative have slightly larger mesogenic core. For all compounds the column cross-section is made of single molecule.

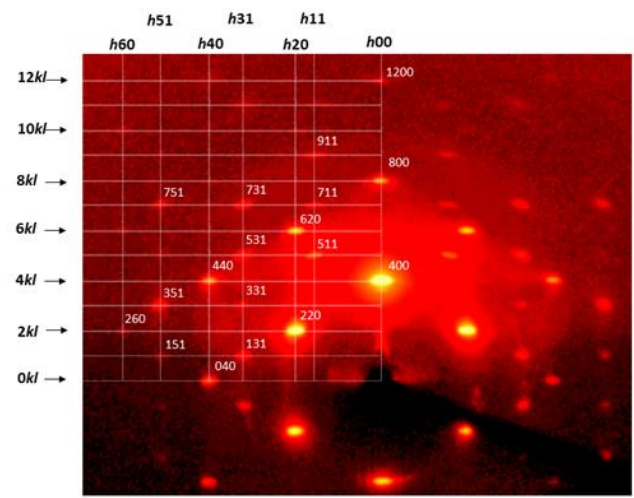

**Figure 3.** XRD pattern of *Fddd* phase of compound **5**. The alignment of sample is induced by surface conditions, domains are orientated with crystallographic axis *b* parallel to the substrate surface.

Compounds **1** and **5** in temperature range below the columnar hexagonal phase, show extraordinary diffraction patterns (*Figure 3*). Apart from six strong signals, at nearly the same position as in *Col_h* phase, multitude of new signals was detected, that could not be indexed assuming any 2D structure. However, the large part of signals fits 2D rectangular lattice with unit cell 2 times enlarged comparing to upper temperature hexagonal phase; such lattice enlargement might be due to presence of additional symmetry elements like 4-fold screw axis or d-type gliding planes. The rest of the signals were assigned to the reflections from the planes involving third direction – apparently in this phase columns are additionally modulated along their axes. Interestingly, the periodicity along the columns is very different for both materials, being three times higher for compound **1** than for compound **5**, ~150 Å and ~50 Å, respectively.

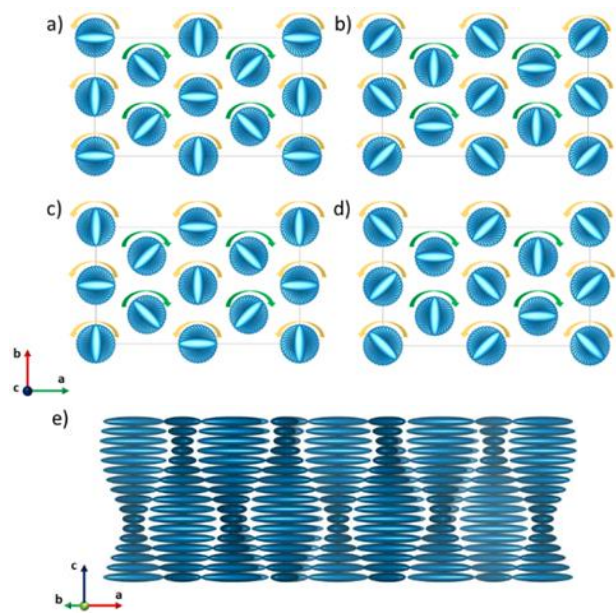

**Figure 4.** Model of the structure of columnar *Fddd* phase built form helical columns with anisotropic cross sections, the arrangement of columns in the crystallographic unit cell in **ab** plane at different **z**-levels: a) **z**=0, b) **z**=0.25, c) **z**=0.5, d) **z**=0.75 and in **ac** plane (e). Green and yellow arrow represent right and left handed columns, respectively.

The reflection extinction conditions (only hkl signals with: h+k, l+h, k+l = 4n and hk0 signals with h+k=2n were present) point to *Fddd* symmetry. However, it should be noted that while for compound **5** a perfect fit of the XRD pattern is possible assuming *Fddd* symmetry, in the pattern of compound **1** there are few signals that are forbidden by this symmetry. The presence of these signals may indicate the local disorder and large number of defects. If the additional electron density in third orthogonal direction comes from helical arrangement of molecules along columns, the symmetry elements of *Fddd* group (gliding planes) generate in crystallographic cell 8 columns, four left handed and four right handed. Moreover, if cross-sections of the columns is elliptical, symmetry elements will produce in each xy plane four different orientations of ellipsoids (*Figure 4a-d*). This means that neighbouring columns are mutually shifted along z-direction; such a shift assures better pack ing condition of columns accommodating the lath-like objects, that twist along the column (*Figure 4e*). At high diffraction angle range only diffused signal centred at the ~4.5Å was observed in *Fddd* phase, evidencing liquid crystalline character of the phase. Taking the mean distance between molecules in the columns, the average twist between neighbouring molecules in column is 10 and 30 deg for materials **1** and **5**, respectively. The signal positions and lattice parameters for *Fddd* phase for compound **1** and **5** were collected in *Table 3* and in *Table S2*. It should be pointed out that the columnar phase with *Fddd* symmetry has been recently described for different type of mesogenic molecules, in case of materials studied by G. Ungar's group[26] the column cross section was made of 2-3 molecules forming flexible aggregates.

Molecules studied here contain π-conjugated bond system and thus are expected to exhibit strong luminescent properties in visible range. The data related to absorption and emission properties are collected in *Table 4*. The UV–Vis spectra obtained in dichloromethane (DCM) solution in most cases show single maximum in visible range (400-500 nm), only compound **3** (and also non-liquid crystalline compounds **6** and **8**, see SI) have additional vibrational structure of absorption band. The origin of the absorption peaks is due to π-π* transitions and possibly a small charge transfer component from substituents to quinoxaline unit[29]. It was found that the larger conjugated π bond system is, the more the absorption maximum is red-shifted.

**Table 4.** Luminescent properties of compounds **1-5**, measured in DCM solution and crystal, $\phi$ stands for quantum yield; $\lambda_{exc}^{DCM}$ is the absorption (excitation) wavelength in DCM, $\lambda_{em}^{DCM}$ and $\lambda_{em}^{s.st.}$ is the emission wavelength in DCM and solid state respectively

| No. | $\lambda_{exc}^{DCM}$ /nm | $\lambda_{em}^{DCM}$ /nm | $\lambda_{em}^{s.st.}$ /nm | Stokes shift /nm | $\phi$ solution | $\phi$ crystal |
|---|---|---|---|---|---|---|
| 1 | 470 | 545 | 555 | 75 | 0,06 | 0,16 |
| 2 | 490 | - | - | - | - | - |
| 3 | 508 | - | - | - | - | - |
| 4 | 407 | 492 | 506 | 85 | 0,25 | 0,17 |
| 5 | 439 | 580 | 536 | 141 | 0,02 | 0,07 |

Regarding fluorescence, two of the compounds (**2, 3**) did not respond to the light excitation at the absorption band. Other materials exhibited fluorescence both in solution (in DCM solvent) and in solid crystal (slightly shifted, *Table 3*); with rather large Stokes shift, the largest value, nearly 150 nm, was found for compound **5**. The large Stokes shift in solution might indicate the strong excitation-induced molecular geometry changes or formation of intramolecular charge transfer state. However, the change of solvent polarity (also toluene solution was tested) affected only weakly the position of absorption and emission maxima (~5 nm). The measured fluorescence efficiency, $\phi$, in solution and solid state correlates with the flexibility of the molecules. The significant quantum efficiency was found only for compound **4** (~25%), with the most flexible molecular structure. Apparently, the conformational freedom and non-planar molecular geometry prevented strong interaction of molecules in solid state and in local aggregates formed in solution. On the other hand non detectable fluorescence for compounds **2** and **3**, with flat and rigid molecular structure might suggest formation of strong H-aggregates, it is well known that luminescence is quenched for co-facially stacked molecules [30].

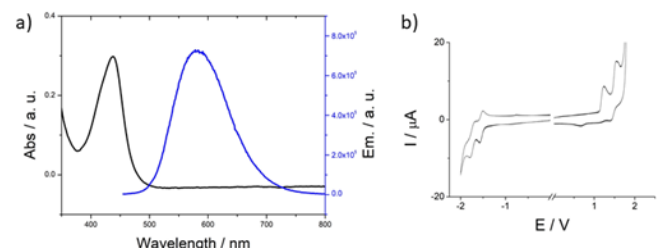

**Figure 5.** a) Absorption and emission spectra for compound **5** and b) cyclic voltammogram for compound **4**, measured in dichloromethane and scaled vs ferrocene potential.

In order to determine electron-donating/accepting properties of studied materials the electrochemical measurements were performed using cyclic voltammetry (CV) method for all compounds except for the compound **3** that was not sufficiently soluble at room temperature (*Table 4*). At both, negative and positive potentials voltammograms exhibit two peaks, each peak representing the reversible mono-electron process.

**Table 5.** The $E_{red}/E_{ox}$ are reduction and oxidation potentials in regard to ferrocene (Fc/Fc$^+$) taken from cyclic voltammetry. The values of the formal potentials, $E_{1/2}$, obtained for each redox step in the voltammograms.

| No. | $E_{red}$ [V] | $E_{ox}$ [V] | $E_{LUMO}$ [eV] | $E_{HOMO}$ [eV] | $\Delta E$ [eV] |
|---|---|---|---|---|---|
| 1 | -1,004 | 1,181 | -3,272 | -5,495 | 2,223 |
| 2 | -0,861 | 1,132 | -3,454 | -5,445 | 1,881 |
| 3 | | non soluble at room temperatures | | | |
| 4 | -1,474 | 1,139 | -2,841 | -5,454 | 2,613 |
| 5 | -1,354 | 1,248 | -2,961 | -5,563 | 2,602 |

*$E_{1/2}$ was approximated as $(E_{pa} + E_{pc})/2$, where $E_{pa}$ and $E_{pc}$ are the oxidation and reduction peak potentials, respectively. $E_{LUMO}$ and $E_{HOMO}$ are calculated energies of LUMO and HOMO levels, respectively. $\Delta E$ is the energy gap between those levels.*

While the positions of positive potential signals that were related to oxidation of alkoxy groups at the outer phenyl rings were nearly the same for all studied compounds, the positions of negative potential signals differ significantly depending on the structure of the mesogenic core. The processes at negative potential are related to reversible reduction of quinoxaline unit, they take place at smallest potentials for material **2** with the most enlarged π-bond system. The energy difference between HOMO and LUMO states, determined form the difference between first redox signals at positive and negative potentials, was ~ 2,5 eV for all compounds except for material **2,** for which it was considerably smaller, ~1.8 eV.

Because of bi-polar, i.e. electron-donating/accepting character of the compounds, they were further tested by ToF method to determine the charge mobility along columns [6]. All LC materials were studied but the photocurrent was registered only for compounds **2** (with smallest energy band gap taken from CV) and **3**. For these compounds the absorption onset was detected at the longest wavelength (539 nm and 543 nm).

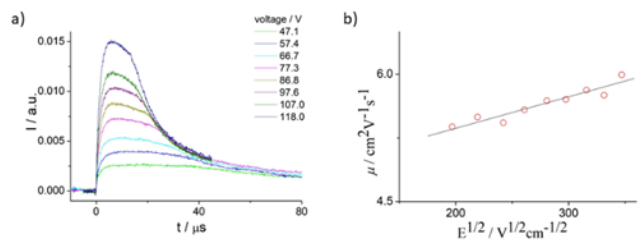

**Figure 6.** ToF experiment data for compound **2**; a) ToF transients current of holes for *Col$_h$* phase at constant temperature (150 °C) for different values of applied electric field; b) field dependence of hole mobility (Poole-Frenkel relation, $\ln\mu \sim E^{1/2}$[31]) in *Col$_h$* phase.

The hole and electron currents were registered for compound **3** after illumination of the sample with UV light pulse (355 nm), however their polydispersive character prevented determination of the charge mobility quantitatively. For the compound **2,** the hole photocurrent with clear cut-off vs. time was registered (*Figure 6a*), which enabled determination of the charge mobility. In measured voltage range μ was $5,38 – 6,00 \cdot 10^{-4}\ cm^2/V \cdot s$ (Figure 6b), the charge mobility $\mu_0$ extrapolated to zero voltage, $4,66 \cdot 10^{-4}\ cm^2/V \cdot s$, was considerably lower than reported record-high values in LC phases[32], but rather typical for Col$_h$ phase[33,34]. Additionally, apart from UV pulse also excitation with visible light was tested (532 nm). The hole and electron current signals were significantly smaller than upon illumination with UV light (*Figure S6*).

In summary, among studied quinoxaline derivatives those having flexible molecular structure tend to form not only simple columns arranged into hexagonal lattice, but also helical columns, in which molecules twist along the column; such chiral columns are arranged into complex 'antichiral' structure with *Fddd* symmetry. The symmetry of the phase results from frustration induced by tendency to arrange the equal amount of left- and right-handed columns into triangular lattice. In resulting structure the helical columns are vertically shifted to ensure better packing conditions of elliptical objects forming helices, at each level of crystallographic cell. The molecular flexibility affects also other material properties. For molecules in which the peripheral phenyl rings in the mesogenic core are free to rotate, the non-planarity of the structure prevents the aggregation-caused fluorescence quenching. On the other hand, flat molecular structure with extended π-conjugated system quenches the fluorescence, apparently the excited state relaxes through formation of excitons rather than through the photon emission. The excitons relatively easy dissociate leading to strong hole current along columns observed for materials with stiff or semiflexible structure of the core. For one of the materials bi-polar charge transfer was found upon excitation with UV or visible light.

## ASSOCIATED CONTENT

**Supporting Information**. This material is available free of charge via the Internet at http://pubs.acs.org.

## AUTHOR INFORMATION

**Corresponding Author**

* prybak@chem.uw.edu.pl

**Author Contributions**

The manuscript was written through contributions of all authors.

**Funding Sources**

The work was supported by the project: NCN Maestro 2016/22/A/ST5/00319

## ACKNOWLEDGMENT

This work is dedicated to the memory of Prof. Michael Hanack.

# Complex structure of chiral columns made of achiral quinoxaline derivatives with semi-flexible cores

Paulina Rybak*, Adam Krowczynski, Jadwiga Szydlowska, Damian Pociecha and Ewa Gorecka

**Abstract:** Novel mesogenic materials, quinoxaline derivatives with semi-flexible cores were synthesized and studied. Some of these materials apart of Col$_h$ (columnar hexagonal) phase form new type of 3D columnar structure with larger crystallographic unit cell and Fddd symmetry. Structure is result of contradictory requirments packing of columns of opposite chirality into hexagonal lattice. The fluorescent properties could be easily tuned by modification of molecular structure, compounds with the extended pi conjugated systems form aggregates and florescence is quenched, for molecules with flexible structure in which molecular parts are free to rotate the fluorescence quantum yield reaches 25%. Material for which fluorescence is suppressed show strong hole photocurrent. The suitability of new materials for preparing organic field transistors was tested.

**Table of Contents**

**Experimental Procedures**

**Synthesis**

**Scheme S1.** Schematic representation of the synthesis of obtained compounds.

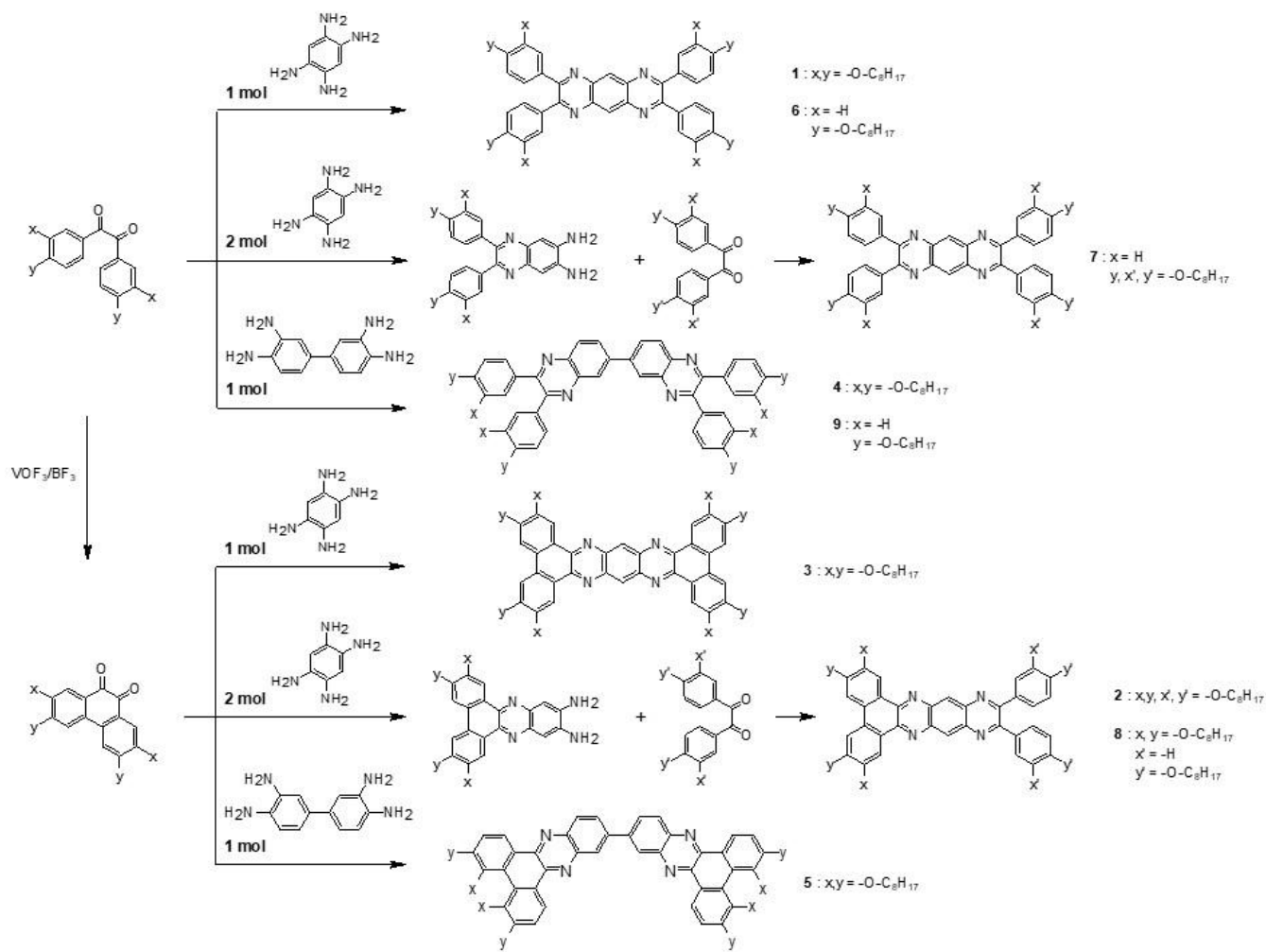

Preparation of compounds: pyrazino[2,3-g]quinoxaline, 2,3,7,8-tetrakis[3,4-bis(octyloxy)phenyl] (**1**), pyrazino[2,3-g]quinoxaline, 2,3,7,8-tetrakis[4-(octyloxy)phenyl] (**6**) and pyrazine [2,3-g]quinoxaline, 2,3-di[3,4-bis(octyloxy)phenyl], 7,8-di[4-(octyloxy)phenyl] (**7**) followed the prescription given by Walsh [1], Wang [2] and Vishwakarma [3]. Phenantrene-quinone was prepared after Kato [4]. After coupling the same procedure as in abovementioned prescriptions was followed to obtained compounds [2,3,6,7,13,14,17,18]-octa-(octyloxy)-tetrabenzo-[a,c,l,n]-[9,11,20,22]-tetraaza-pentacene (**3**), [2,3,6,7]-tetra-(octyloxy)-[12,13]-(di-([3,4]-di-octyloxy)-phenyl)-[9,11,14,16]-tetraaza-tetracene **2** and [2,3,6,7]-tetra-(octyloxy)-[12,13]-(di-4-octyloxy-phenyl)-[9,11,14,16]-tetraaza-tetracene **8**; The intermediate product (coupled and uncoupled diamines) in synthesis of **7**, **2** and **8** was prepared according to the prescription described by Krzyczkowska et al. [5].

Synthesis with 3,3'-diaminobenzidine of compounds: 2,3,2',3'-tetrakis-(3,4-bis-octyloxy-phenyl)-[6,6']biquionoxaline (**4**), 11,11'-bi(di-[2,3,6,7,2',3',6',7']-octa-(octyloxy)-benzo)-[a,c]-phenazine (**5**) and 2,3,2',3'-tetrakis-(4-octyloxy-phenyl)-[6,6']-biquionoxaline **9** was performed in the same conditions as in the case of benzene-1,2,4,5-tetraamine; solvents used for attaching the dibenzoyls to the amines were acetic acid and isopropanol.

**Elemental analysis**

**NMR and mass spectra:**

[1]H and [13]C NMR spectra were recorded on Agilent 400 MHz and Bruker 500 MHz NMR spectrometers. Mass spectroscopy was conducted on Micromass LCT.

**1:** $^1$H NMR (400 MHz, CDCl$_3$): δ = 0.82-1.87 (m, 120 H); 3.86 (t, J=6.7 Hz, 8 H); 4.02 (t, J=6.7 Hz, 8 H); 6.86 (d, J= 8.3 Hz, 4 H); 7.15 (d, J=2.0 Hz, 4 H); 7.2 (dd, J=2.0 Hz, 8.3 Hz, 4 H); 8.89 (s, 2 H)
$^{13}$C NMR (400 MHz, CDCl$_3$): δ= 154.68; 150.31; 148.68; 140.18; 131.53; 127.89; 123.09; 115.30; 113.00; 69.19; 69.16; 31.84; 31.83; 29.40; 29.37; 29.33; 29.28; 29.20; 29.13; 29.02; 26.02; 22.69; 22.67; 14.09
HRMS exact mass calculated for C$_{98}$H$_{150}$N$_4$O$_8$: 1512.2616; found: 1512.0453

**2:** $^1$H NMR (400 MHz, CDCl$_3$): δ = 0.87-2.04 (m, 120 H); 3.88 (t, J=6.7 Hz, 4 H); 4.04 (t, J=6.7 Hz, 4H); 4.28 (t, J=6.6 Hz, 4 H); 4.37 (t, J=6.6 Hz, 4 H); 6.88 (d, j=8.5 Hz, 2 H); 7.19 (d, J=2.0 Hz, 2 H); 7.2 (dd, J=2.0 Hz, 8.5 Hz, 2 H); 7.67 (s, 2 H); 8.82 (s, 2 H); 9.1 (s, 2 H)
$^{13}$C NMR (400 MHz, CDCl$_3$): δ= 154.40; 152.35; 150.29; 149.42; 148.67; 143.76; 140.75; 139.66; 131.73; 127.58; 127.12; 123.61; 123.16; 115.34; 112.93; 109.40; 106.52; 69.47; 69.19; 31.88; 31.85; 31.84; 29.48; 29.42; 29.39; 29.36; 29.34; 29.33; 29.31; 29.29; 29.23; 29.17; 26.19; 26.15; 26.04; 22.71; 22.69; 22.67; 14.10
HRMS exact mass calculated for C$_{98}$H$_{148}$N$_4$O$_8$: 1510.2457; found: 1510.2963

**3:** $^1$H NMR (500 MHz, CDCl$_3$, 50°C): δ = 0.92-2.03 (m, 120 H); 4.18 (t, J=6.6 Hz, 8 H); 4.30 (t, J=6.6 Hz, 8 H); 7.45 (s, 4 H); 8.63 (s, 4 H); 8.96 (s, 4 H)
$^{13}$C NMR (500 MHz, CDCl$_3$, 50°C): δ= 152.23; 149.45; 143.36; 140.02; 126.94; 123.9; 109.64; 106.72; 69.61; 69.17; 31.99; 31.95; 29.68; 29.63; 29.44; 29.42; 26.36; 26.31; 22.76; 22.73; 14.10; 14.07
HRMS exact mass calculated for C$_{98}$H$_{146}$N$_4$O$_8$: 1508.2298; found: 1508.2406

**4:** $^1$H NMR (400 MHz, CDCl$_3$): δ = 0.84-1.87 (m, 120 H); 3.82-3.88 (m, 8 H); 4.01 (broad t, J=6.7 Hz, 8 H); 6.85 (broad d, J=8.4 Hz, 4 H); 7.1 (d, J=2.1 Hz, 2 H); 7.12 (d, J= 2.1 Hz, 2 H); 7.13-7.17 (m, 4 H); 8.18 (dd, J=1.9 Hz, 8.7 Hz, 2 H); 8.25 (d, J=8.7 Hz, 2 H); 8.54 (d, J=1.9 Hz, 2 H)
$^{13}$C NMR (400 MHz, CDCl$_3$): δ= 153.79; 153.37; 149.94; 148.70; 141.20; 140.82; 140.66; 131.71; 131.69; 129.63; 128.99; 127.16; 122.85; 115.30; 115.28; 113.16; 113.12; 69.16; 31.84; 31.83; 29.41; 29.38; 29.33; 29.28; 29.23; 29.15; 26.03; 22.69; 22.67; 14.09
HRMS exact mass calculated for C$_{104}$H$_{154}$N$_4$O$_8$: 1588.3576; found: 1588.3443

**5:** $^1$H NMR (500 MHz, CDCl$_3$, 50°C): δ = 0.89-1.99 (m, 120 H); 4.24-4.27 (m, 8 H); 4.31-4.36 (m, 8 H); 7.63 (s, 2 H); 7.64 (s, 2 H); 8.29 (dd, J=2.0 Hz, 8.8 Hz, 2 H); 8.37 (d, J=8.8 Hz, 2 H); 8.72 (broad s, 4 H)
$^{13}$C NMR (500 MHz, CDCl$_3$, 50°C): δ= 151.93; 149.58; 142.38; 141.96; 141.74; 141.27; 140.25; 129.75; 128.48; 127.33; 126.64; 126.60; 124.01; 123.88; 109.12; 109.07; 106.80; 106.73; 69.79; 69.25; 31.96; 31.94; 29.63; 29.60; 29.57; 29.56; 29.42; 29.41; 26.36; 26.33; 26.30; 22.75; 22.73; 14.10; 14.08
HRMS exact mass calculated for C$_{72}$H$_{90}$N$_4$O$_4$: 1075.5094; found: 1076.0352

**6:** $^1$H NMR (400 MHz, CDCl$_3$): δ = 0.88-1.84 (m, 60 H); 4.0 (t, J=6.6 Hz, 8 H); 6.89 and 7.58 (AA', BB', J=8.8 Hz, 16 H); 8.87 (s, 2H)
$^{13}$C NMR (400 MHz, CDCl$_3$): δ= 160.20; 154.51; 140.21; 131.44; 131.27; 127.86; 114.30; 68.10; 31.82; 29.37; 29.24; 29.22; 26.04; 22.66; 14.11
HRMS exact mass calculated for C$_{66}$H$_{86}$N$_4$O$_4$: 999.4134; found: 999.4204

**7:** $^1$H NMR (400 MHz, CDCl$_3$): δ = 0.87-1.87 (m, 90 H); 3.86 (t, J=6.7 Hz, 4 H); 4.0 (t, J=6.6 Hz, 4H); 4.02 (t, J=6.6 Hz, 4 H); 6.87 (d, j=8.3 Hz, 2 H); 7.89 and 7.58 (AA', BB', J=8.8 Hz, 8 H); 7.15 (d, J=2.0 Hz, 2 H); 7.21 (dd, J=2.1 Hz, 8.3 Hz, 2 H); 8.88 (s, 2 H)
$^{13}$C NMR (400 MHz, CDCl$_3$): δ= 160.22; 154.65; 154.58; 150.30; 148.68; 140.25; 140.15; 132.19; 131.55; 131.43; 131.25; 127.88; 123.08; 121.61; 115.31; 114.32; 114.04; 113.88; 113.02; 69.20; 29.17; 68.22; 68.11; 31.85; 31.83; 31.81; 29.40; 29.37; 29.36; 29.33; 29.31; 29.28; 29.23; 29.21; 29.13; 26.03; 22.69; 22.67; 22.65; 14.09
HRMS exact mass calculated for C$_{82}$H$_{118}$N$_4$O$_6$: 1255.8375; found: 1256.0212

**8:** $^1$H NMR (400 MHz, CDCl$_3$): δ = 0.87-2.03 (m, 90 H); 4.01 (t, J=6.6 Hz, 4 H); 4.22 (t, J=6.6 Hz, 4 H); 4.35 (t, J=6.6 Hz, 4 H); 6.9 and 7.59 (AA', BB', J=8.8 Hz, 8 H); 7.57 (s, 2 H); 8.73 (s, 2H); 9.01 (s, 2 H)
$^{13}$C NMR (400 MHz, CDCl$_3$): δ= 160.10; 153.97; 152.07; 149.13; 143.38; 140.50; 139.52; 131.52; 131.49; 127.39; 126.78; 123.34; 114.15; 109.13; 105.96; 69.29; 69.01; 68.09; 31.95; 31.91; 31.85; 29.63; 29.59; 29.44; 29.42; 29.40; 29.28; 26.28; 26.20; 26.09; 22.76; 22.73; 22.68; 14.16; 14.13; 14.12
HRMS exact mass calculated for C$_{82}$H$_{116}$N$_4$O$_6$: 1253.8216; found: 1253.7984

**9:** $^1$H NMR (400 MHz, CDCl$_3$): δ = 0.86-1.83 (m, 60 H); 3.98 (broad t , J=6.6 Hz, 8 H); 6.88 and 7.52 (broad AA', BB', J=8.8 Hz, 16 H); 8.15 (dd, J=2.0 Hz, 8.8 Hz, 2 H); 8.23 (d, J=8.8, 2 H); 8.51 (d, J=2.0 Hz, 2 H)
$^{13}$C NMR (400 MHz, CDCl$_3$): δ= 159.85; 159.84; 153.69; 153.25; 141.25; 140.82; 140.71; 131.41; 131.40; 131.24; 129.62; 128.97; 127.14; 114.33; 68.08; 31.81; 29.37; 29.24; 29.23; 26.04; 22.66; 14.10
HRMS exact mass calculated for C$_{72}$H$_{90}$N$_4$O$_4$: 1075.5094; found: 1075.7056

**Identification of mesophases**

**Calorimetry**
Calorimetric studies were performed with a TA DSC Q200 calorimeter, samples of mass from 1 to 3 mg were sealed in aluminium pans and kept in nitrogen atmosphere during measurement, and both heating and cooling scans were performed with a rate of 5–10 K/min.

**Microscopic studies**
Optical studies were performed by using the Zeiss Imager A2m polarizing microscope equipped with Linkam heating stage.

**X-Ray diffraction**

The X-ray diffraction patterns were obtained with the Bruker D8 GADDS system (CuKα line, Goebel mirror, point beam collimator, Vantec2000 area detector). Samples were prepared as droplets on a heated surface.

**Luminescence studies**

Absorption spectra were measured for diluted in methylene chloride solutions (concentration about 2-4·10$^{-2}$ g·dm$^{-3}$) using spectrometer: Shimadzu PC3100. Emission spectra was detected using FluoroLog HORRIBA Jobin Ivon. Fluorescence quantum yields were measured using the standard method, with fluorescein dissolved in 0,1M NaOH as referential material.

$$\varphi = \varphi_{ref} \frac{Grad_x}{Grad_{ref}} \cdot \frac{n_{DCM}^2}{n_{NaOH}^2}$$

(fluorescein quantum yield: $\varphi_{ref} = 0{,}79$; refractive indices: $n_{DCM} = 1{,}424$; $n_{NaOH} = 1{,}335$)

**Charge mobility**

The TOF experiments were performed in a conventional setup. The 10 µm thick cell was used with ITO electrodes covered with homogeneously aligning surfactant, the cells were filled using capillary forces, the applied voltage was in the range of 11–117 V. The transient photocurrent was measured over 5 kΩ and 100 kΩ (form electron current) resistor and recorded with 300 MHz digitizing oscilloscope (Agilent Technologies DSO6034A) triggered by the laser pulse. The estimated response time of the whole setup was less than 2.5 µs. The charges (holes and electrons) were generated by a short light pulse (355 nm and 532 nm wavelengths, ≈8 ns pulse width) coming from a solid-state laser EKSPLA NL202. The sample was illuminated by a single pulse manually triggered to give the sample enough time for relaxation. To reduce a noise the data were collected over 16 runs and averaged. If the registered hole photocurrent curves were nondispersive the clear cutoff enabling precise determination the transient time, τ. The transient time τ was determined as the intersection of two lines tangential to the plateau and "current tail." Form the transient time τ the charge (hole) mobility was calculated according to formula: µ = d/(τE), where d is the sample thickness (cm), E is the strength of electric field (V cm$^{-1}$), and τ is the time of flight (τ/s).

**Electrochemistry**

In order to record electrochemical data we measured cyclic voltammetry (CV) and differential pulse voltammetry (DPV) using the bipotentiostat (*CH* Instruments 750E, Austin, Tx, USA) in registered in DCM solutions with tetrabutyloammnium hexafluoro phosphate, also in DCM, as supporting electrolyte. Measurements were carried out in the three electrode arrangements, with calomel electrode as the reference electrode, platinum foil as the counter and glassy carbon electrode (GCE, BASi, A=0.070 cm2) as the working electrode. The analyzed samples were deoxygenated prior to measurements by purging with argon (99.999 %) for 20 min and then argon was passed over the solution surface. At each measurement series the reference electrode potential was calibrated using ferrocene (Fc/Fc+ ) in the same supporting electrolyte solution and the calibration constant $E_{ferr}$ were found. The formal potentials were determined from cyclic voltammetry waves as averaged oxidation/reduction potential. When CV waves were not clearly shaped the square wave voltammetry was applied. The energies of HOMO/LUMO levels of a given compound ($E_{HOMO}$ and $E_{LUMO}$) were evaluated from its first oxidation and the first reduction potentials. The energy $E_{HOMO}$ and $E_{LUMO}$ levels were calculated according to the following equations $E_{HOMO} = - (E_{ox} - E_{ferr} + 4.8)$ eV and $E_{LUMO} = - (E_{red} - E_{ferr} + 4.8)$ eV.

## Results

The mesogenic compounds were shown and described in the main article. Here we present additional data regarding those mesogens along with other obtained compounds (Fig. S2), that did not exhibit any additional phases between crystallization and isotropic liquid but did give interesting spectral response. Among all obtained compounds, in order to exhibit liquid crystalline properties it is crucial to have two chains substituted in every single peripheral ring.

**Figure S1.** Non-liquid crystalline compounds

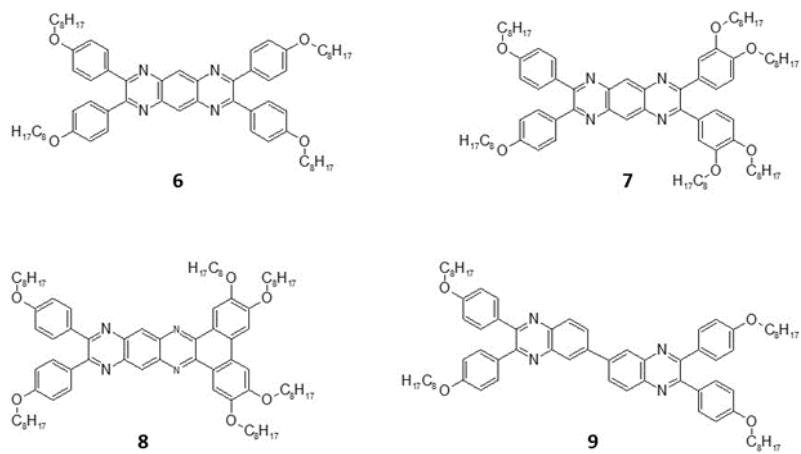

**Table S1.** Phase transition temperatures [°C] and associated thermal effects [J/g] (in parentheses) for obtained compounds;

|   | Phase transition |
|---|---|
| 6 | **Cry** - 192.84 (54.89) - **Iso** |
| 7 | **Cry** - 139.96 (42.15) - **Iso** |
| 8 | **Cry** - 164.06 (47.29) - **Iso** |
| 9 | **Cry** - 126.60 (35.81) - **Iso** |

**Figure S2.** DSC plot of compound **1** and **5**

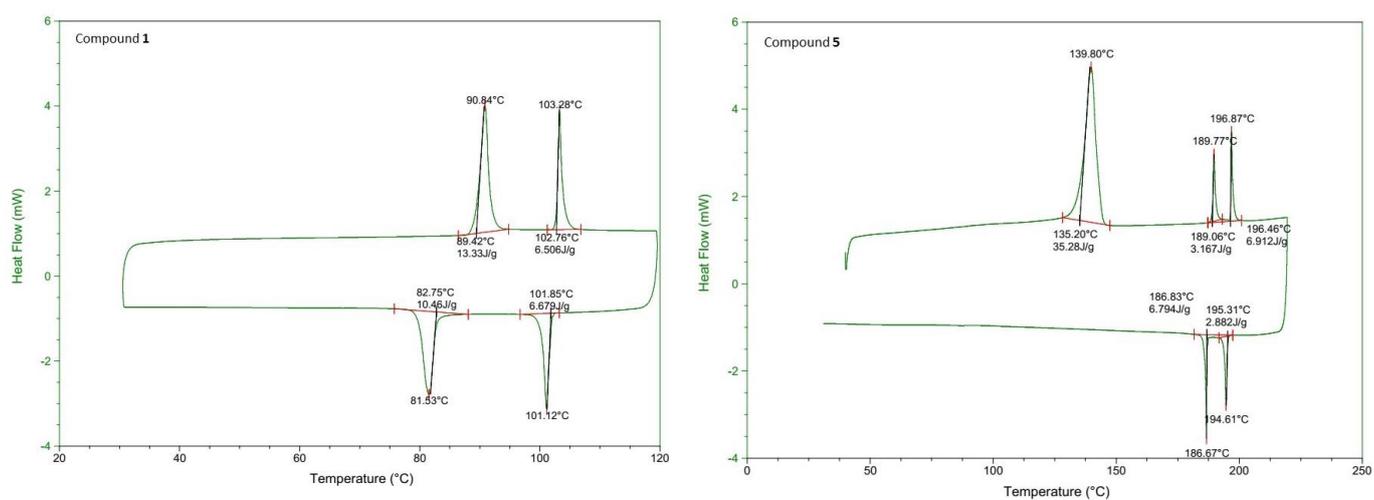

## X-Ray studies

**Figure S3.** XRD pattern representing Col$_h$ phase (a) and Fddd phase (b) for compound **1** at 102°C and 50°C, respectively; c) XRD pattern of Col$_{Fddd}$ phase in compound **1**. Reciprocal lattice for a sample of fiber geometry with fiber axis along a* direction is superimposed.

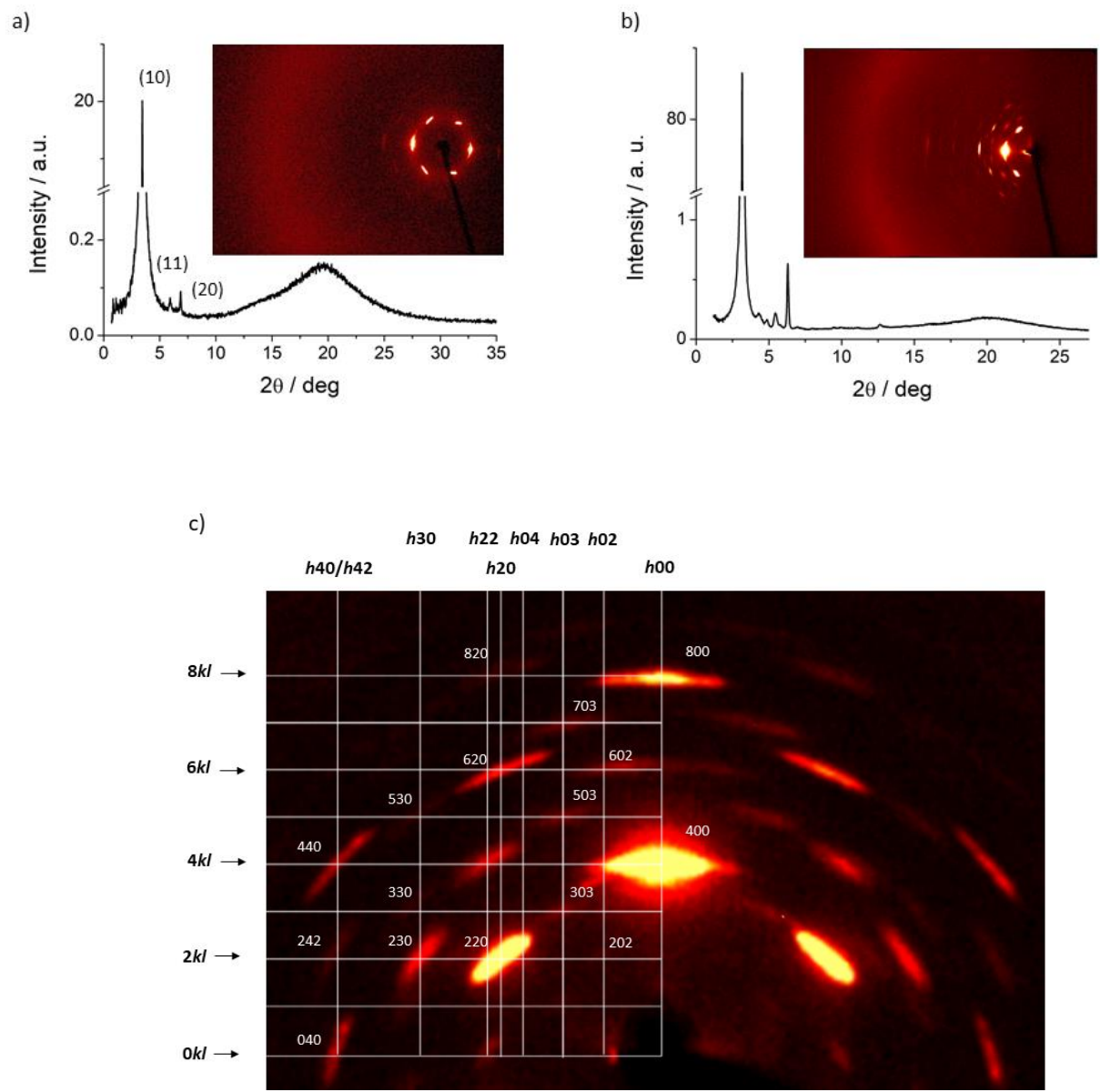

**Table S2.** Experimental and calculated signal positions and lattice parameters for compounds **1** and **5** in *Fddd* phase; calculated d values was according to the given unit cell parameters: for compound **1**: *a*=113.11Å, *b*=65.31Å, *c*=155.42Å and for compound **5**: *a*=107.63Å, *b*=62.14Å, *c*=49.75Å.

| Compound 1 | | | Compound 5 | | |
|---|---|---|---|---|---|
| $d_{exp}$ | $d_{calc}$ | hkl | $d_{exp}$ | $d_{calc}$ | hkl |
| 45.66 | 45.73 | 202 | 27.06 | 26.91 | 220 |
| 30.44 | 31.95 | 303 | 26.90 | 26.91 | 400 |
| 30.31 | 30.10 | 022 | 26.58 | 26.35 | 311 |
| 28.14<br>28.32 | 28.28 | 220<br>400 | 19.16<br>18.82 | 18.83 | 131<br>511 |
| 22.74<br>23.04 | 22.86 | 224<br>404 | 17.04 | 16.88 | 331 |
| 20.56 | 21.18 | 503 | 15.69<br>15.59 | 15.53 | 040<br>620 |
| 20.33 | 20.61 | 422 | 14.36<br>14.30 | 14.30 | 531<br>711 |
| 20.26 | 20.29 | 230 | 13.58<br>13.45 | 13.45 | 440<br>800 |
| 19.42 | 19.76 | 513 | 12.13 | 12.00 | 151 |
| 18.72<br>18.36 | 18.01 | 331<br>317 | 12.03 | 11.98 | 731 |
| 16.16<br>16.25 | 16.33 | 040<br>620 | 11.53<br>11.44 | 11.43 | 351<br>911 |
| 15.42<br>15.47 | 15.61 | 531<br>703 | 10.63 | 10.52 | 551 |
| 15.37 | 15.38 | 242 | 10.25<br>10.26<br>10.15 | 10.17 | 260<br>840<br>10 20 |
| 14.06<br>14.11 | 14.14 | 440<br>800 | 9.55<br>9.46 | 9.49 | 751<br>11 11 |
| 13.26 | 12.93 | 731 | 9.02<br>8.96 | 8.97 | 660<br>12 00 |
| 12.72 | 12.80 | 822 | 8.85<br>8.73 | 8.71 | 171<br>11 31 |
| 12.23 | 12.31 | 824 | 8.16 | 8.10 | 571 |
| 12.22 | 12.30 | 911 | 7.54 | 7.46 | 480 |
| 11.12 | 11.19 | 10 02 | 7.87<br>7.79 | 7.77 | 080<br>12 40 |
| 10.72 | 10.78 | 062 | 7.46<br>7.50 | 7.46 | 14 20<br>10 60 |
| 10.11 | 10.15 | 6 2 12 | | | |
| 10.10 | 10.13 | 11 11 | | | |
| 10.11 | 10.07 | 462 | | | |
| 9.39 | 9.43 | 12 00 | | | |
| 8.94 | 8.99 | 12 22 | | | |

**Table S3.** Absorption and emission peaks, measured in DCM solutions;

Supporting Information

|   | $\lambda_{exc}$ [nm] | $\lambda_{em}$ [nm] | Stokes shift [nm] | $\varphi$ solution | $\varphi$ solid |
|---|---|---|---|---|---|
| 6 | 462 | 502 | 40 | 0.36 | 0.65 |
| 7 | 465 | 543 | 78 | 0.05 | 0.35 |
| 8 | 486 | - | - | - | - |
| 9 | 400 | 446 | 46 | 0.4 | 0.11 |

Compounds **1-3** and **6-8** with the central part of the molecule built with pyrazino-quinoxaline structure, have absorption spectra red – shifted dependent on the extension of the π-conjugated system. Among **1-3** compounds the most red-shifted is mesogen **3**, with the most rigid central core (23 nm shift according to **1**). The same effect can be observed for compounds **4** and **5** (29 nm shift). The other factor that has the influence on the absorption spectra shift is the number of terminal alkoxy chains. This effect is visible comparing spectra of compounds **1** with **6** (25 nm shift), **2** with **8** (19 nm shift) and **4** with **9** (14 nm shift). Those are expected results due to the fact that free electron couple from oxygen can be included in the π-conjugated system. From this two factors bigger impact lays in the rigidity of the core.

When it comes to emission spectra, compounds (**2**, **3** and **8**) with the highest rigidity of the pyrazino-quinoxaline core, where at least two peripheral phenyl rings are coupled, do not show spectral response to the excitation wave. As for the number of terminal chains its impact is not explicit, because compounds **1** and **7** with 8 and 6 alkoxy chains, respectively, have almost identical emission spectra, whereas compound **6**, with 4 alkoxy chains has spectra less red-shifted. For the compounds with the biphenyl-based core (**4**, **5** and **9**) the most red-shifted emission spectra was detected for compound **5** with adjacent phenyl rings and 8 terminal chains.

**Figure S4.** Normalized absorption (a) and emission (b) spectra for all obtained compounds;

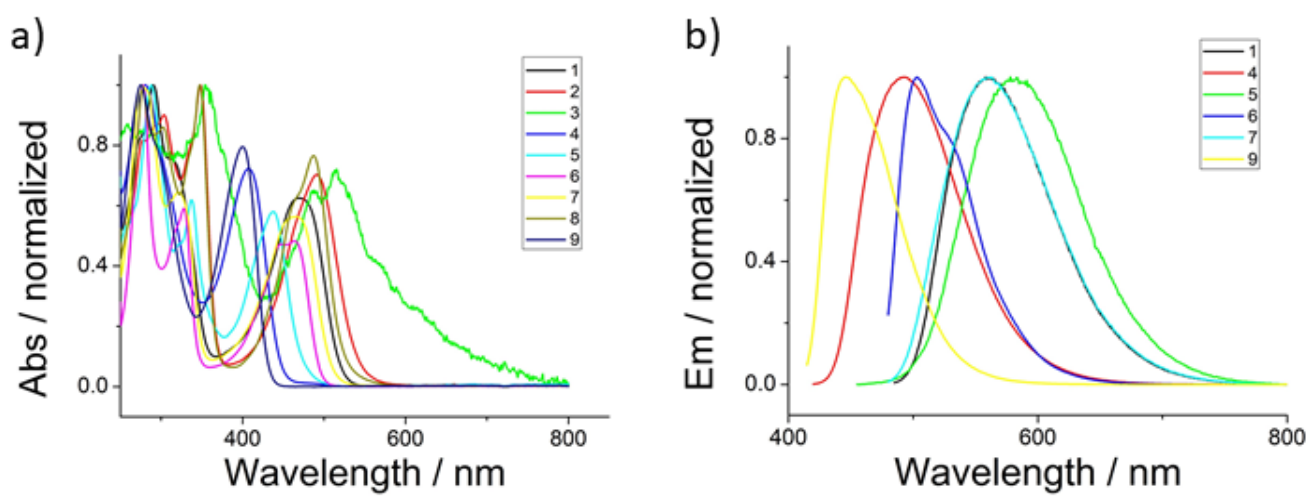

**Figure S5.** Cyclic voltammograms, measured in dichloromethane and scaled vs ferrocene potential.

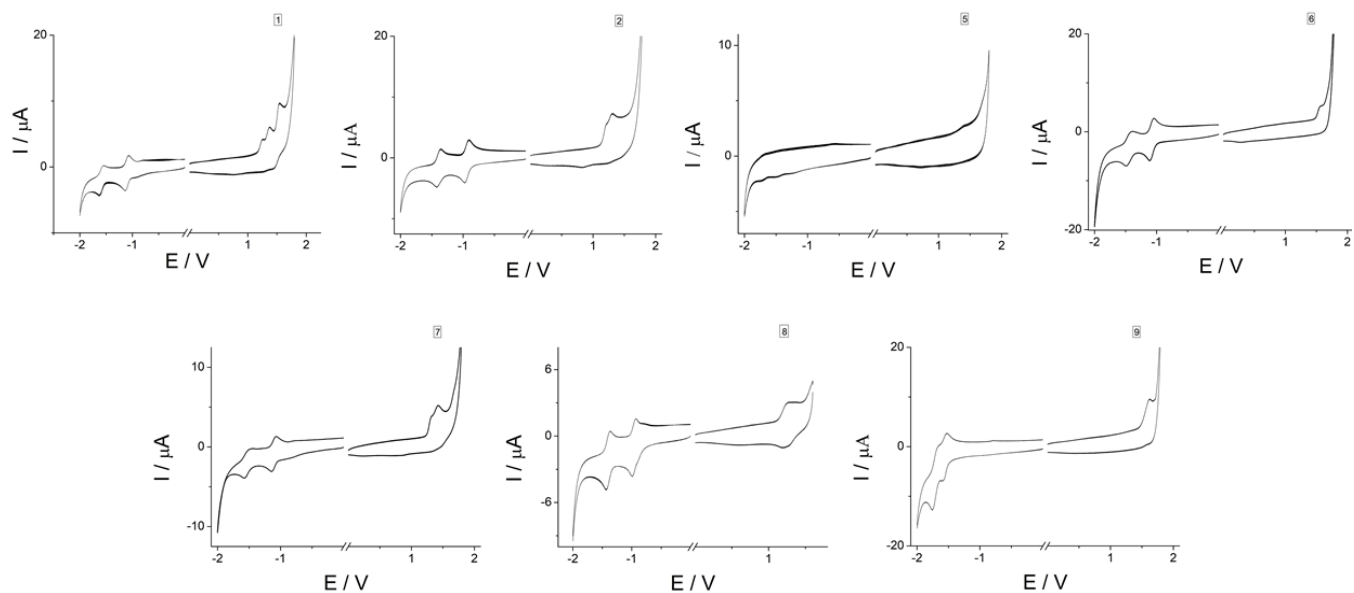

**Table S4.** The $E_{red}/E_{ox}$ are reduction and oxidation potentials in regard to ferrocene (Fc/Fc$^+$) taken from cyclic voltammetry. The values of the formal potentials, $E_{1/2}$, obtained for each redox step in the voltammograms; $E_{1/2}$ was approximated as $(E_{pa} + E_{pc})/2$, where $E_{pa}$ and $E_{pc}$ are the oxidation and reduction peak potentials, respectively. $E_{LUMO}$ and $E_{HOMO}$ are calculated energies of LUMO and HOMO levels, respectively. $\Delta E$ is the energy gap between those levels.

|   | $E_{red}$ [V] | $E_{ox}$ [V] | $E_{LUMO}$ [eV] | $E_{HOMO}$ [eV] | $\Delta E$ [eV] |
|---|---|---|---|---|---|
| 6 | -1.008 | 1.485 | -3.307 | -5.8 | 2.493 |
| 7 | -1.053 | 1.248 | -3.262 | -5.563 | 2.301 |
| 8 | -0.783 | 1.125 | -3.532 | -5.44 | 1.908 |
| 9 | -1.489 | 1.47 | -2.826 | -5.785 | 2.959 |

**Table S5.** Collected electrochemical data compared to the spectral response of all compounds with their molecular structures;

| No. | Formula | Abs. onset / nm, eV | Em. peak / nm, eV | $E_{LUMO}$ /eV | $E_{HOMO}$ /eV | $\Delta E_{LUMO-HOMO}$ / eV |
|---|---|---|---|---|---|---|
| 1 | (structure: tetraaryl-pyrazinoquinoxaline with four 3,4-bis(octyloxy)phenyl substituents) | 520 / 2.38 | 560 / 2.21 | -3,272 | -5,495 | 2,223 |

| # | Structure | | | | | |
|---|---|---|---|---|---|---|
| 2 | (structure) | 539 2.30 | - | -3,454 | -5,445 | 1,881 |
| 3 | (structure) | 543 2.28 | - | - | - | - |
| 4 | (structure) | 444 2.79 | 493 2.51 | -2,841 | -5,454 | 2,613 |
| 5 | (structure) | 473 2.62 | 582 2.13 | -2,961 | -5,563 | 2,602 |
| 6 | (structure) | 495 2.50 | 503 536 2.31 | -3,307 | -5,8 | 2,493 |
| 7 | (structure) | 510 2.43 | 561 2.21 | -3,262 | -5,563 | 2,301 |
| 8 | (structure) | 520<br>2.38 | 527<br>672 Weak<br>2.35<br>1.84 | -3,532 | -5,44 | 1,908 |
| 9 | (structure) | 430<br>2,88 | 446<br>2.78 | -2,826 | -5,785 | 2,965 |

**Figure S6.** ToF results for compound **2**; Green light (532 nm) excitation for holes (a) and electrons (b);

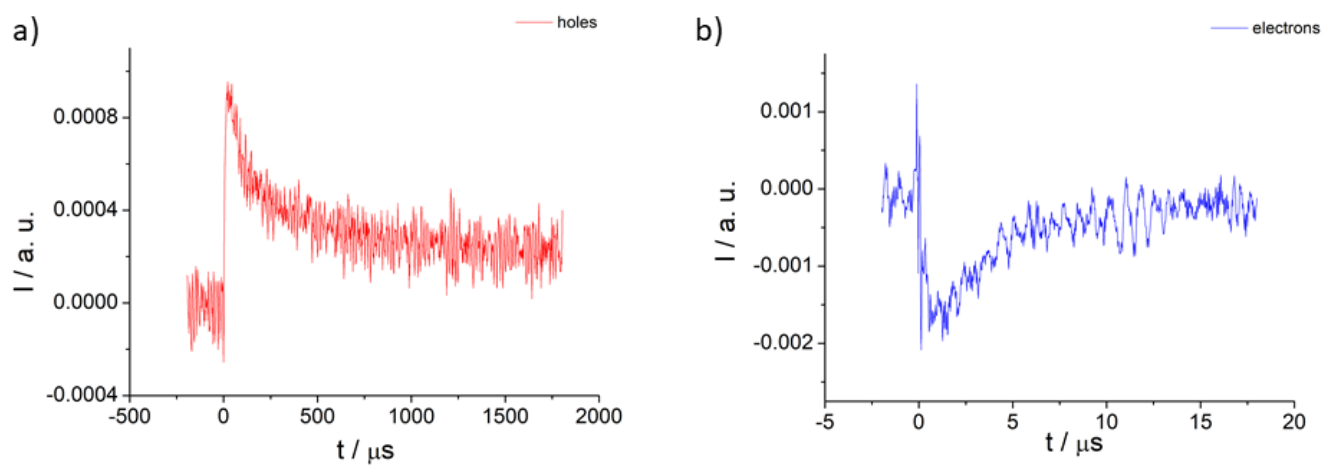

## References

(1) Walsh, C. J.; Mandal, B. K.  A Novel Method for the Peripheral Modification of Phthalocyanines. Synthesis and Third-Order Nonlinear Optical Absorption of β-Tetrakis(2,3,4,5,6-pentaphenylbenzene)phthalocyanine  *Chem Mater* **2000**, *12*, 287.
(2) Wang, X.; Zhou, Y.; Lei, T.; Hu, N.; Chen, E. Q.; Pei, J. A.  Structural-Property Relationship in Pyrazino[2,3-g]quinoxaline Derivatives: Morphology, Photophysical, and Waveguide Properties  *Chem Mater* **2010**, *22*, 3735.
(3) Vishwakarma, V. K.; Nath, S.; Gupta, M.; Dubey, D. K.; Swayamprabha, S. S.; Jou, J. H.; Pal, S. K.; Sudhakar, A. A.  Room-Temperature Columnar Liquid Crystalline Materials Based on Pyrazino[2,3-g]quinoxaline for Bright Green Organic Light-Emitting Diodes  *Acs Appl Electron Ma* **2019**, *1*, 1959.
(4) Kato, S.; Takahashi N Fau - Tanaka, H.; Tanaka H Fau - Kobayashi, A.; Kobayashi A Fau - Yoshihara, T.; Yoshihara T Fau - Tobita, S.; Tobita S Fau - Yamanobe, T.; Yamanobe T Fau - Uehara, H.; Uehara H Fau - Nakamura, Y.; Nakamura, Y.  Tetraalkoxyphenanthrene-fused dehydroannulenes: synthesis, self-assembly, and electronic, optical, and electrochemical properties.
(5) Krzyczkowska, P.; Krówczyński, A.; Kowalewska, J.; Romiszewski, J.; Pociecha, D.; Szydłowska, J.  Mesogenic Enaminoketone Ni(II) Complexes of Phenazine and Quinoxaline Derivatives  *Mol Cryst Liq Cryst* **2012**, *558*, 93.

## Author Contributions

P.R. and A.K. performed  the organic synthesis of reported compounds. P.R. carried out experiments regarding spectroscopy and electrochemistry and along with J.Sz. performed ToF studies. The X-Ray studies were performed and analyzed by P.R., D.P. and E.G.. All authors contributed to the manuscript writing.